# PROGRAMMABLE MOLECULAR COMPOSITES OF TANDEM PROTEINS WITH GRAPHENE-OXIDE FOR EFFICIENT BIMORPH ACTUATORS

Mert Vural[1,2], Yu Lei[3], Abdon Pena-Francesch[2], Huihun Jung[2], Benjamin Allen[4,5], Mauricio Terrones[1,3*], Melik C. Demirel[1,2,4*]

[1] Center for Research on Advanced Fiber Technologies (CRAFT), Materials Research Institute, [2] Department of Engineering Science and Mechanics, [3] Department of Physics, Department of Chemistry, Department of Materials Science & Engineering and Center for 2-Dimensional and Layered Materials, [4] Huck Institutes of Life Sciences, and [5] Department of Biochemistry and Molecular Biology, Pennsylvania State University, University Park, PA, 16802

* Corresponding authors: Melik C. Demirel, Tel: 814 863-2270, mdemirel@engr.psu.edu
Mauricio Terrones, 814 865-0343, mut11@psu.edu

**Abstract**

The rapid expansion in the spectrum of two-dimensional (2D) materials has driven the efforts of research on the fabrication of 2D composites and heterostructures. Highly ordered structure of 2D materials provides an excellent platform for controlling the ultimate structure and properties of the composite material with precision. However, limited control over the structure of the adherent material and its interactions with highly ordered 2D materials results in defective composites with inferior performance. Here, we demonstrate the successful synthesis, integration, and characterization of hybrid 2D material systems consisting of tandem repeat (TR) proteins inspired by squid ring teeth and graphene oxide (GO). The TR protein layer acts as a unique programmable molecular spacer between GO layers. As an application, we further demonstrate thermal actuation using bimorph molecular composite films. Bimorph actuators made of molecular composite films (GO/TR) can lead to energy efficiencies 18 times higher than regular bimorph actuators consisting of a GO layer and a TR protein layer (i.e., conventional bulk composite of GO and TR). Additionally, molecular composite bimorph actuators can reach curvature values as high as 1.2 $cm^{-1}$ by using TR proteins with higher molecular weight, which is 3 times higher than conventional GO and TR composites.

# 1. Introduction

Graphene-based materials exhibit unique physical properties including mechanical strength and thermal and electrical conductivity. [1] These qualities have enabled the synthesis of novel hybrid nanomaterials [2] with applications in artificial nacre structures, flexible electronics, energy storage systems and mechanical actuators. [3-7] Despite their immense potential in various fields of engineering, graphene-based composites fail to sustain these properties in bulk form, since it is extremely difficult to control their microstructure. In order to synthesize graphene-based materials in bulk form with controllable/programmable order and microstructure, [8] researchers have proposed various approaches to material synthesis and fabrication. [9, 10] The most common approaches include mixing an organic matrix to control graphene oxide (GO) arrangement and interlayer spacing (e.g., molecular composites), and layer-by-layer assembly of GO and organic macromolecules into heterostructures. [11-13]

It is noteworthy that molecular composites offer an easier and faster method for the fabrication of graphene-based composites in bulk form via solution-processed self-assembly. However, it is challenging to control the desired molecular order throughout the entire material for molecular composites, as the material order is simply governed by molecular interactions between GO and the organic matrix. [11, 14, 15] In this context, layer-by-layer assembly provides a better alternative regarding the molecular order in the composites, since each layer is deposited individually. [16, 17] Despite the precise material order in composites prepared using layer-by-layer assembly, the rate of deposition is extremely low, which generates an almost insurmountable bottleneck for bulk material fabrication. [12] In summary, it is essential to establish novel organic matrices that offer control over the molecular interactions with GO for fabricating programmable bulk molecular composites. Composites consisting of polymers with the ability to hydrogen bond with graphene oxide have been studied as a common approach to this problem. [18] Previous reports describe the synthesis of molecular composites of GO [19] and hydrogen bonded polymers (e.g., PMMA[11], PVA[20]). However, the polymers used in these studies lacked the potential for supramolecular chemistry, which is essential for single-parameter control of microstructural features, such as GO-layer interspacing. Biopolymers such as structural proteins present a fine alternative for this application, as the resulting microstructure is highly dependent on the specifics of the hydrogen bonds they can establish. This can enable control over molecular interactions and ultimately the structure of the organic matrix in these composites by adjusting a single parameter (e.g. molecular weight, concentration, and crystal size).

For practical applications, these molecular composites can initiate strong mechanical actuations with the help of the bimorph structures. The implementation of these structures can lead to significantly greater energy-efficiency of actuation, low threshold voltage for actuation, and enhanced ultimate curvature, due to the unique molecular nature of these layers. Specifically, repetitive proteins can provide precise control of molecular morphology at the nanoscale to build bio-inspired molecular composite systems. Here, we report the synthesis and fabrication of 2D-molecular composites, consisting of layers of GO and layers of semi-crystalline organic materials consisting of self-assembling tandem repeat (TR) proteins. We assembled composites of TR and GO into bimorph thermal actuators and evaluated their performance. These novel molecular composite bimorph actuators can facilitate thermal actuation at voltages as low as ~2 V (which is independent of actuator shape and size), and they boast energy efficiencies 18 times better than regular bimorph actuators assembled using bulk GO and TR films. Moreover, it is possible to further improve the extent of deformation of these composites by implementing a TR protein with higher molecular weight to form molecular composites. These actuators can reach curvature values as high as 1.2 cm$^{-1}$, which is unattainable for regular bimorph thermal actuators of GO.

## 2. Experimental

### 2.1 Protein expression of Tandem Repeat (TR) proteins

Tandem repeat proteins was prepared following protocol reported by Jung *et al*. [21] Protein expression is initiated by a single colony, which was inoculated and grown overnight in 5mL of LB with ampicillin (100 μg/mL). The overnight culture were scaled up to 2 L (i.e., four 500mL LB media) and was grown on a shaker at 210 rpm and 37°C for 5 hours. When the cultures reached OD600 of 0.7 - 0.9, IPTG was added to the final concentration of 1 mM and shaking was continued at 37°C for 4 hours. Then, the cells were pelleted at 12,000 rpm for 15 minutes and stored at -80°C. After thawing cell pellets were resuspended in 300 mL of lysis buffer (50 mM Tris pH 7.4, 200 mM NaCl, 1 mM PMSF, and 2mM EDTA), and lysed using a high-pressure homogenizer. The lysate was pelleted at 14,000 rpm for 1 hour at 4°C. The lysed pellet was washed twice with 100 mL of urea extraction buffer (100 mM Tris pH 7.4, 5 mM EDTA, 2 M Urea, 2% (v/v) Triton X-100), and then washed with 100 mL of washing buffer (100mM Tris pH 7.4, 5 mM EDTA). Sample collection in both washing steps (urea extraction and final wash) was performed by centrifugation at 5000 rpm for 15 minutes. The resulting recombinant-protein pellet was dried with a lyophilizer (Labconco, FreeZone 6 plus) for 12 h. The final yield of expressed protein was approximately 15 mg/1 L of bacterial culture.

## 2.2  Graphene Oxide Synthesis and Purification

Graphene oxide was prepared following protocol reported by Marcano *et al.* [22] Briefly, 5 g of graphene was dispersed in a mixture of 200 mL of $H_2SO_4$ and 40 mL of $H_3PO_4$, and then 25 g of potassium permanganate was slowly added. After reaching 40 °C with 1 h of oxidation, the graphene exfoliated, and the mixture is stirred for 3.5 h of oxidation resulting in a dark brown mixture. The solution is poured slowly in a mixture of 600 mL of cold water with 40 mL of 35% $H_2O_2$ and left overnight to allow for complete neutralization of potassium permanganate. The supernatant of the yellow GO dispersion was decanted, and the solid GO cake was dispersed in 1 L of 5 wt % $H_2SO_4$. Following centrifugation of the GO dispersion at 4000 rpm for 5 min, the supernatant was decanted. This step was repeated three times with distilled water. While the acid is washed, the GO expands and the third time the GO pellet shows a two-layer structure. Finally, GO was concentrated into a slurry with pH 3.0–3.5 and 0.9% of solids by centrifugation at 6000 rpm for 3 h.

## 2.3  Composite fabrication

The TR proteins (TR15, TR25, TR42) were dissolved in dimethylsulfoxide (DMSO) to prepare 20 ml solution with a protein concentration of 7.5 mg/ml. Freeze-dried GO flakes were dissolved in DMSO to prepare 30 ml solution with a GO concentration of 1 mg/ml. GO solution is placed in a sonic bath for an hour to individually disperse GO flakes in DMSO. The GO solution is then added to the protein solution dropwise. The mixture is placed into a bath sonicator for 30 minutes prior to flocculation of the composite films. Molecular composite film was made by filtration of the resulting mixture through an Anodisc membrane filter (d: 47mm, 0.2 mm pore size; Whatman), followed by vacuum drying and peeling from the filter.

## 2.4 Actuator fabrication

Rectangular samples (0.8mm, 35mm) were cut from circular composite films. TR42 synthetic proteins were dissolved in Hexafluoro-2-propanol (HFIP) to prepare 1 ml solution with a protein concentration of 50 mg/ml. TR42 solution was cast on the rectangular samples and the samples were dried 30 minutes for solvent evaporation. Solid polydimethylsiloxane masks were placed on the film prior to gold film deposition using sputtering. A 60nm thick gold layer was deposited using sputter deposition.

## 2.5 Material Characterization

The scanning electron microscopy images were acquired using FEI Nova NanoSEM 630 with an accelerating voltage of 5 keV. The EDS patterns were acquired using built in EDS detector of the FEI Nova NanoSEM 630 system. High resolution transmission electron microscopy (HRTEM) were performed using a FEI Talos TEM at an accelerating voltage of 200 kV. X-ray diffraction experiments were performed using reflection mode with PANalytical XPert Pro MPD (Cu$K\alpha$ radiation, $\lambda=1.5406$Å, operating at 40 keV, cathode current of 20 mA) under standard laboratory conditions. Fourier transform infrared spectroscopy (Thermo Scientific Nicolet 6700 FT-IR) data were acquired using an attenuated total reflection mode, and Happ-Genzel apodization with 4 cm$^{-1}$ resolution from 400 to 4000 cm$^{-1}$. For each spectrum 256 scans were performed. Gaussian curve fitting was performed as described elsewhere. [21] Thermogravimetric analysis (TGA) was performed on a TA instruments Q50 coupled to a Pfeiffer vacuum mass spectrometer. For each measurement samples were initially left for equilibration for 30 minutes. After equilibration samples are heated from 25º C to 1000º C at a heating rate of 5º C/min under helium gas flow (90 ml/min).

## 2.6 Actuation measurements

The experiments were performed under constant voltage, which was supplied through the actuator samples using a Keithley Sourcemeter 2400. The current was simultaneously measured during actuation experiments using the same device under different voltage conditions. The curvature of actuators was calculated from the images acquired during the actuation with digital camera using image analysis software ImageJ. The temperatures of the samples were monitored using an embedded thermocouple and infrared thermometer during thermal actuation.

## 3. Results and Discussion

Squid ring teeth (SRT) proteins are H-bonded thermoplastic elastomers extracted from the tentacles of the squid suction cups (Figure 1A) that exhibit a reversible transition from a solid to a melt and, therefore, can be thermally shaped into any 3D geometry (e.g. fibers, colloids, and thin films). [23] These proteins have recently been shown to have excellent mechanical properties in both wet and dry conditions, [24, 25] exceeding most natural and synthetic polymers while having the unique capability of self-healing [26, 27] and self-assembly. [21] SRT proteins, which have a segmented copolymer sequence consisting of amorphous and crystalline domains, have been investigated as alternatives to synthetic polymers. [21] These unique structural proteins exhibit such extraordinary physical properties due to a unique semi-crystalline structure. In light of these studies, it has been recently demonstrated that it is possible to control and improve the physical properties (i.e., mechanical and thermal) of SRT-based biopolymers using a specific polypeptide sequence for both crystalline and amorphous regions. [21] Heterologous expression of these tandem repeat proteins in bacteria has enabled bulk synthesis of these unique biopolymers (Figure 1B). Genes for these tandem-repeat SRT proteins are synthesized via a modified rolling-circle-amplification method. [21] The number of tandem repeat (TR) units defines the ultimate material properties of the protein by controlling the molecular weight and interconnectivity of the crystals in these synthetic proteins. TR proteins with 4, 7, and 11 repeat units were prepared, corresponding to molecular weights of 15.6 (TR15), 25.7 (TR25), and 40.5 kDa (TR42), respectively, as shown in the protein gel of Figure 1C.

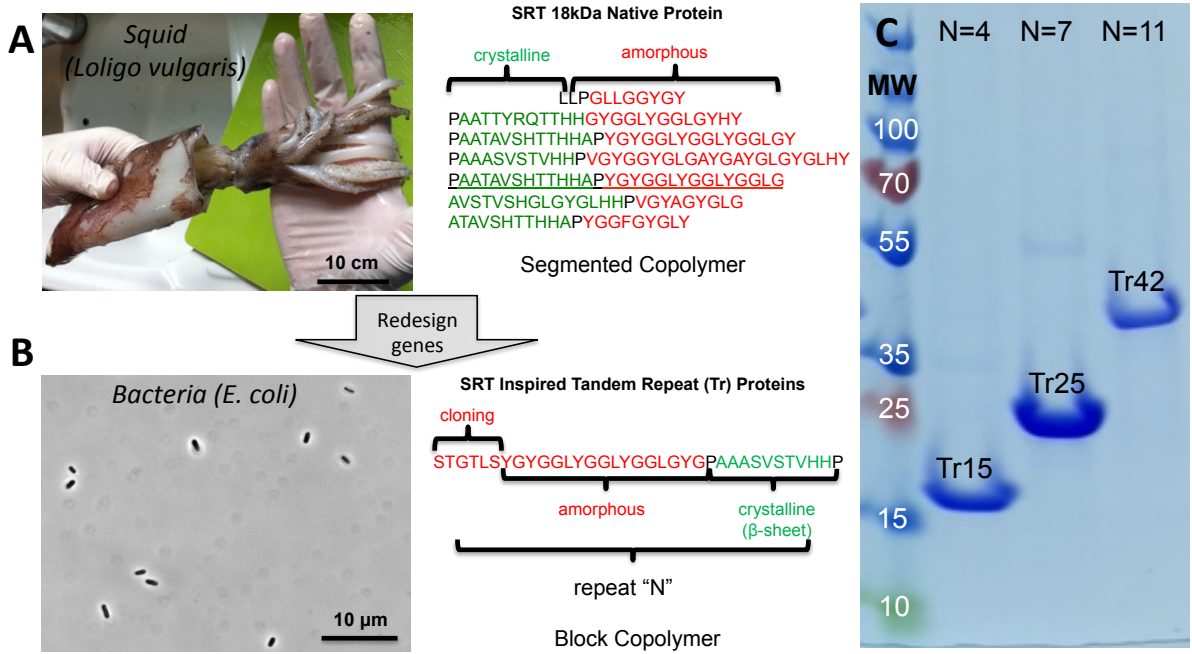

**Figure 1** A) Native Squid ring teeth (SRT) proteins are found in the tentacles of the squid suction cups. However, these proteins show random segmented copolymer sequence with varying crystalline and amorphous domains. B) Engineered sequences of SRT inspired tandem repeat proteins, expressed in *E. coli*, have ordered block copolymer morphology, which is ideal for controlling interlayer distances of 2d-layered materials. C) SDS/Page gel chromatography identifying molecular weight values for three tandem repeat proteins (i.e., Tr15, Tr25, and Tr42) with specific repetitions of 4,7, and 11 respectively.

Note that proteins provide monodisperse and exact molecular weight, which is a unique attribute in obtaining precise control of nanoscale morphology compared to their polydispersed synthetic polymer counterparts. The crystal dimensions of these repetitive TR proteins are relatively independent of the number of repeat units. [21] The crystal domains of these TR proteins consist of anti-parallel β-sheets formed by 4 strands of the crystalline polypeptide sequence. This β-sheet orientation results in a crystal 3 nm long in the polypeptide-backbone direction and 2 nm wide in the hydrogen-bonding direction. The amorphous sequence separates these crystals with tie chains corresponding to separation of 3 nm. [21] The interconnectivity of these tie chains is highly dependent on the number repeat units. In addition, the crystals of these TR proteins do not demonstrate stacking of multiple β-sheets (i.e., a major advantage of TR proteins compared to other structural proteins such as silk), which makes them an ideal candidate for generating, intercalated composites with 2D materials. We combined these TR proteins with large area (2500 μm$^2$) single sheet GO (thickness of 8.5 Å with 1:1 C/O ratio) to generate molecular composites that assemble into specific structures. The interlayer distance of these structures can be controlled using a single parameter (i.e., the number of repetitions in the sequence of the TR proteins). This provides nanoscale control on the structure of the composite, and consequently on the material properties.

The fabrication of free-standing molecular composite films of TR proteins and GO were performed using a vacuum-assisted self-assembly (VASA) method. [11] This technique enables assembly of highly ordered composites of 2D materials in bulk form. TR proteins and GO were dissolved in a common organic solvent (dimethyl sulfoxide, DMSO). These solutions were mixed and homogenized using ultrasonication. The resulting homogeneous solution was filtered directionally through anodized aluminum oxide membranes using a vacuum-assisted solvent filtration apparatus (Figure 2A). The rate of the assembly process was controlled by the vacuum, which led to highly ordered composites with alternating layers of protein and 2D materials. After complete removal of the solvent, the resulting freestanding molecular composite films were peeled from the inorganic membrane (Figure 2B).

Prior to experimental characterization, composite films were dried in a vacuum chamber to remove excess solvent molecules. The initial characterization of the resulting composites was performed using electron microscopy, which demonstrated a compact stacking of alternating GO and protein layers (Figure 2C, Figure S1). Although the scanning electron microscopy (SEM) image shows several surface defects, the cross-section of the freestanding composite film appears void-free (Figure 2C). The proper alternating stacking of the GO and TR protein layers can also be observed in the high-resolution transmission microscopy (HRTEM) image provided in the inset of Figure 2C. Complementary to structural analysis by electron microscopy, we performed chemical analysis of the composite films using energy-dispersive X-ray spectroscopy (EDS). The EDS analysis has demonstrated a clear outline of the material distribution for GO and TR protein. The images corresponding to carbon and oxygen signals showed patterns covering the entire cross-section of the composite film homogeneously, as these elements are common for both GO and TR protein (Figure 2D (ii, iii)). On the other hand, the image corresponding to the nitrogen signal gave a discontinuous pattern of stacked lines, which is expected since nitrogen should only be found in the TR protein layers (Figure 2D(iv)).

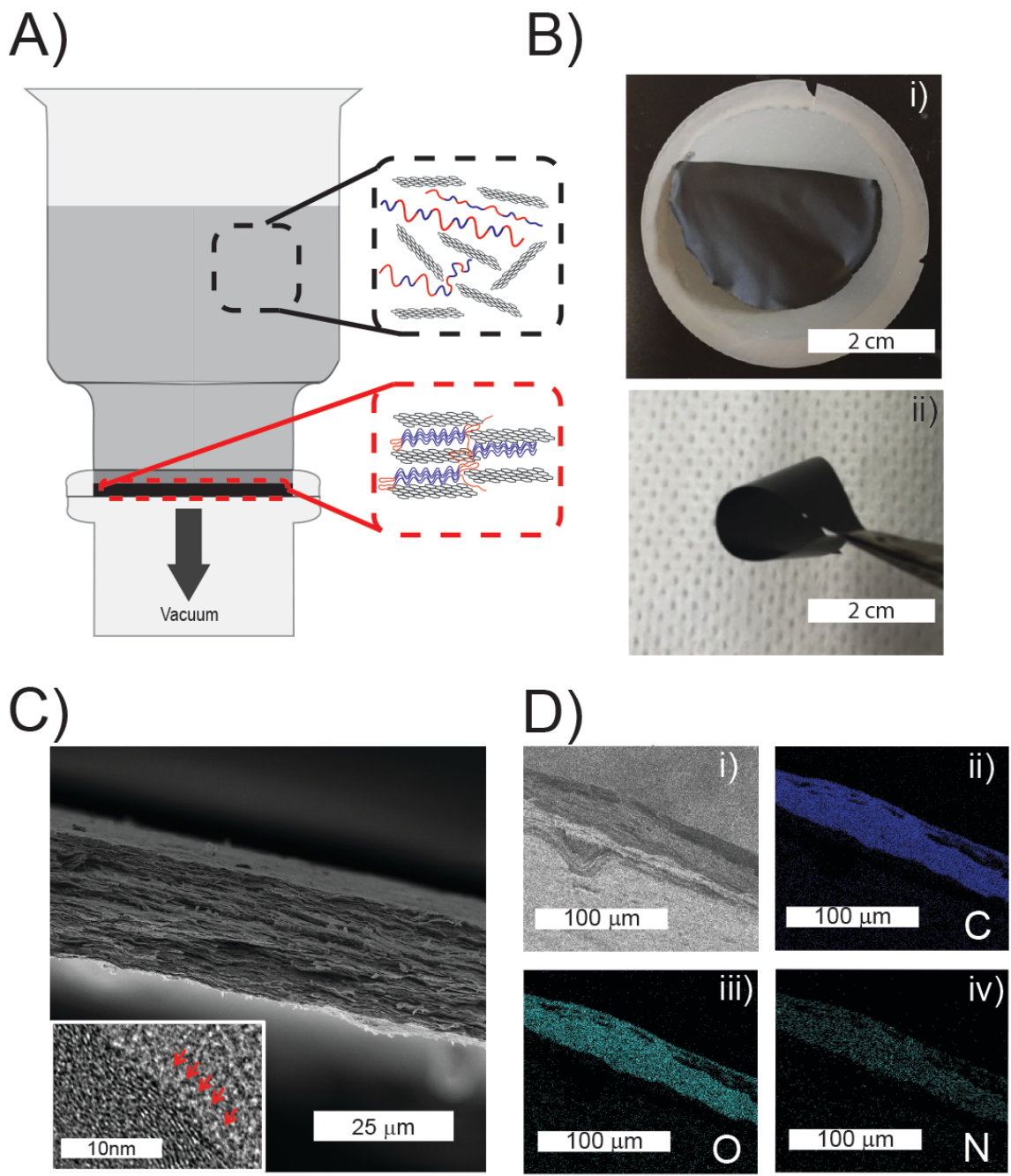

**Figure 2.** A) Schematic illustration of vacuum assisted self-assembly (VASA) of 2D molecular composites. B) Image of free-standing molecular composite consisting of GO and TR protein with 42 kDa molecular weight. C) Cross-section scanning electron microscope (SEM) and transmission electron microscope (inset) image of molecular composite consisting of GO and TR protein with 25 kDa molecular weight. D) i) Backscattered electron image, and energy dispersive X-ray spectroscopy (EDS) patterns of ii) carbon, iii) oxygen, iv) nitrogen for molecular composite consisting of GO and TR protein with 25 kDa molecular weight.

To study the microstructure of the molecular composites further, X-ray diffraction (XRD) analysis was performed on molecular composites GO/TR15, GO/TR25, GO/TR42 (Figure 3A). The XRD characterization demonstrated that only the GO/TR42 composite film resulted in diffraction peaks consistent with both GO ((001), 2θ=9.7°) and tandem protein ((100), 2θ=6.8° (200), 2θ=19.4°). In contrast to GO/TR42, XRD analysis of GO/TR15 and GO/TR25 only showed the diffraction peak corresponding to the (001) plane of GO. This peak visibly shifted toward higher angles for composites consisting of proteins with higher molecular weight (Figure 3B). This indicates the interlayer distance between GO layers increases with increasing the molecular weight of the TR proteins. The protein TR15 kDa increased the interlayer distance of 2D materials by 4.8 Å (Figure S2). On the other hand, synthetic protein TR42 increased the interlayer distance of 2D materials by 9.6 Å (Figure S2). This increase in distance between GO sheets corresponds to the approximate size of a single β-sheet layer. This separation dimension along with the crystalline protein peaks observed in the XRD data of GO/TR42 suggests that these composite films have single β-sheets separating GO sheets. In order to clarify the role of β−sheets in this structure, we performed Fourier transform infrared spectroscopy (FTIR) (Figure 3C). The infrared absorbance spectrum (400-4000 cm$^{-1}$) of each composite demonstrates characteristic peaks of both GO (1084 cm$^{-1}$ (C-O), 1620 cm$^{-1}$ (C=C), 1728 cm$^{-1}$ (C=O)) and TR proteins (Amide I (1580-1720 cm$^{-1}$), Amide II (1460-1570 cm$^{-1}$). The analysis of β-sheet content was performed by inspecting the amide I region of the spectrum (Figure 3D), and by deconvoluting absorbance peaks corresponding to GO and each secondary structure of the TR proteins (Figure S3). The Amide I region of the infrared spectrum of GO/TR15 and GO/TR25 showed relatively lower absorbance values at wavenumbers ranging between 1600 and 1640 cm$^{-1}$, yet GO/TR42 demonstrated the strongest absorbance in this spectral region, which corresponds to absorbance of β-sheets (Figure 3D). The deconvolution of each absorbance peak revealed that the area of β-sheet peaks for GO/TR15 and GO/TR25 is much lower than the area of β-sheet peaks for GO/TR42 (Figure S3). Consequently, the relative protein crystallinity in GO/TR15 (18 %) and GO/25 (27%) is much lower than the protein crystallinity in GO/TR42 (36%) (Figure S3). These values support the XRD characterization; when cast on their own, these proteins demonstrate similar crystallinity values (35%-45%). [21] In order to identify the orientation of β-sheets in GO/TR15, powder diffraction studies were performed. The powder diffraction data clearly demonstrates the existence of β-sheets, which assemble potentially in gallery regions in between planar GO stacks (i.e., hidden to XRD analysis of composite films due to periodicity in stacking direction), instead of separating GO layers as in GO/TR42 composites (Figure S4).

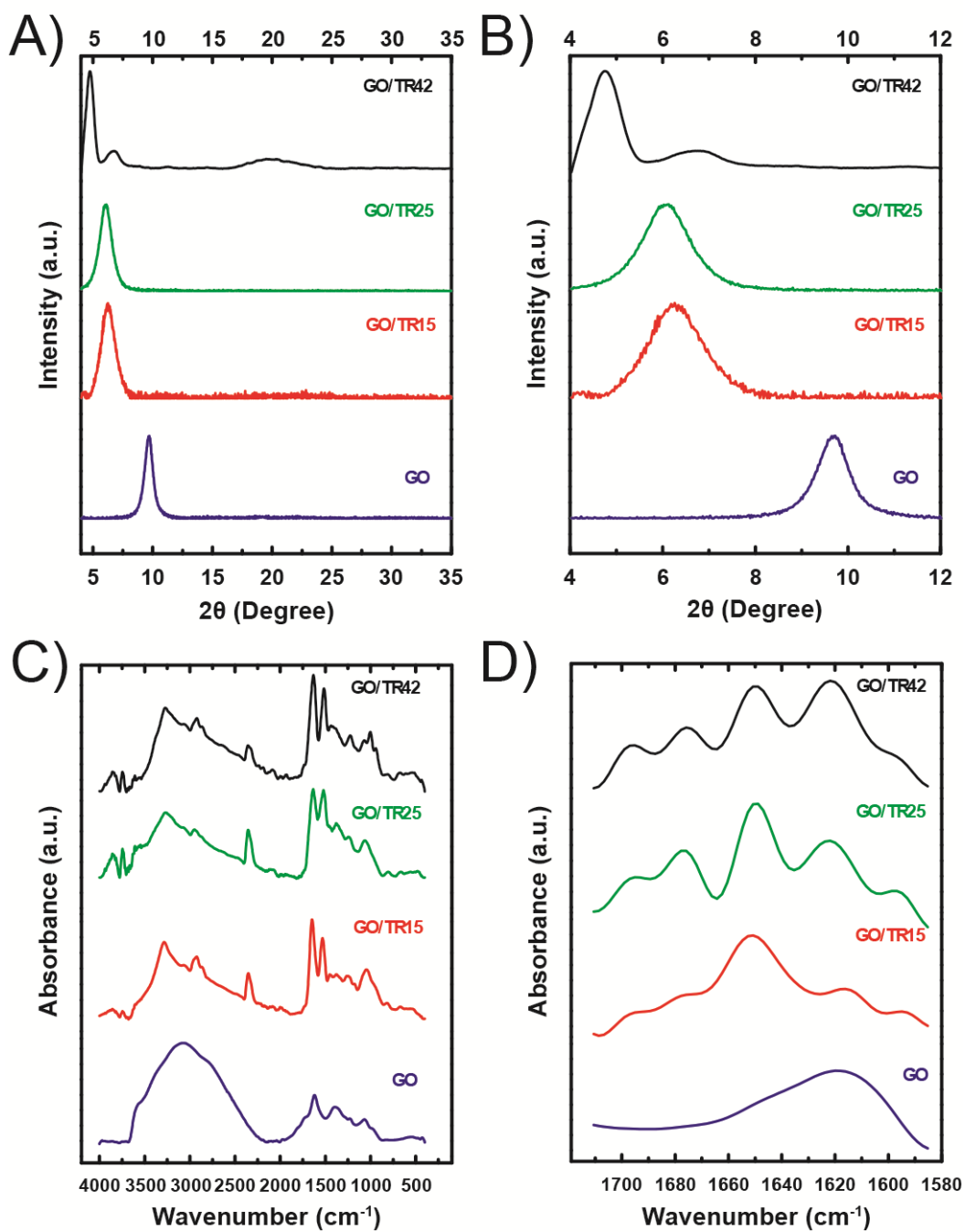

**Figure 3.** A) XRD spectra for GO, and molecular composites prepared using TR15, TR25, TR42 proteins. B) XRD spectra of GO, and molecular composites prepared using TR15, TR25, TR42 proteins focused GO (001) plane. C) FTIR spectra of GO, and molecular composites prepared using TR15, TR25, TR42 proteins. D) Amide-I band region (1720-1580 cm$^{-1}$) of the deconvoluted FTIR spectra for GO, and molecular composites prepared using TR15, TR25, TR42 proteins.

To further study the origin of the increase in interlayer distance for GO, we performed thermogravimetric analysis coupled with mass spectrometry (TGA-MS). This characterization is essential to identify the composition of these materials, since the observed differences in interlayer distance could originate from different protein contents in each sample. TGA-MS studies showed that molecular composites prepared with TR proteins of different molecular weights show very similar trends for mass loss with increasing temperature (Figure 4A). The first derivative of TGA data is used to identify the origin of mass loss in molecular composites for composition analysis (Figure S5). The composition analysis performed on the TGA-MS data revealed that these composites have very similar protein contents (55 wt% ±10 wt%) (Figure 4B). This is a clear indication that the variation in interlayer distance originates from the structures of the TR proteins and their interactions with GO; the protein layer acts as a programmable molecular spacer between GO layers. To support the chemical composition studies (TGA, EDS), we performed mass spectroscopy analysis on these composites, which confirmed the existence and the amount of the protein in these molecular composite films (Figure S5).

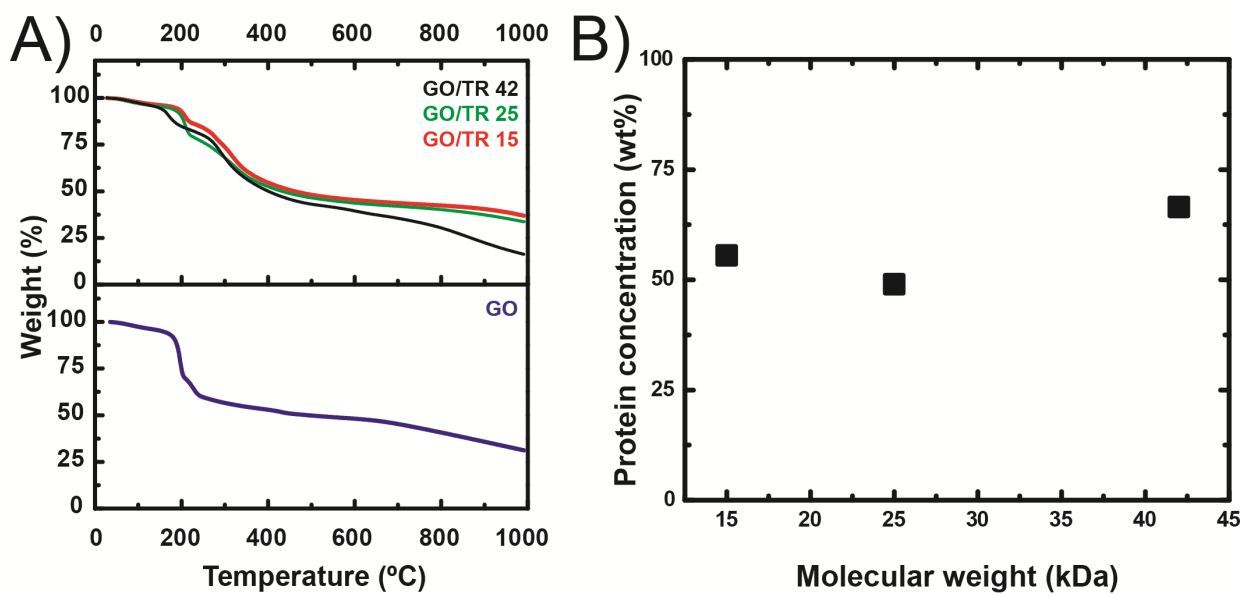

**Figure 4.** A) Thermogravimetric analysis coupled with a mass spectrometry (TGA-MS) for GO, and molecular composites consisting of GO/TR15, GO/TR25 and GO/TR42 show approximately 55% protein content, which agrees with the density measurements (i.e., $\rho$(Go/TR15) = 1.82, $\rho$(Go/TR25) = 1.78, and $\rho$(Go/TR42) = 1.61) of these composites. B) Protein concentration in molecular composites as a function of the molecular weight of the proteins.

Following the structural and compositional characterization, we employed molecular composites to develop thermal actuators, which combine the excellent thermal conductivity of GO (GO:300 W/mK and SRT:0.3 W/mK) with the superior thermal expansion coefficient of TR proteins (SRT: $-95\times10^{-6}$ K$^{-1}$, and GO: $-50\times10^{-6}$ K$^{-1}$). [19, 28] To highlight the impact of the molecular composites on the performance of thermal actuators, we fabricated two distinct types of bimorph composites: regular bimorph GO actuators and molecular composite bimorph GO actuators (see Table S1). Regular bimorph GO actuators consist of a flocculated GO film (thickness, t = 30 μm) and a TR protein film (TR42, t = 30 μm) to initiate thermal actuation and support the GO film. Molecular composite bimorph GO actuators are assembled using a flocculated molecular composite film (t = 40 μm) and a TR protein film (TR42, t = 30 μm). The composite film is responsible for homogeneous and rapid heat dissipation and uniform actuation. The GO layers in the composite dissipate heat properly and the protein layers initiate thermal expansion, triggering actuation due to their intrinsically superior thermal expansion coefficient. The second film of TR42 protein is implemented to improve the ultimate curvature of the actuator. The thermal actuation is initiated using joule heating.

The heater electrodes are patterned on the films using gold sputtering and shadow masking. The thickness of the gold electrodes is measured as 60 nm, which is significantly thinner than GO (30 μm), molecular composite (t = 40 μm) and TR protein films (t = 20 μm). This is important for minimizing the influence of the electrodes on thermal actuation. The fabricated actuators are tested under fixed voltage conditions to investigate the corresponding curvature occurring due to thermal actuation (Figure 5B, C). Regular bimorph actuators can only reach 0.12 cm$^{-1}$ curvature values at relatively low power values (0.2 W/cm$^2$) (Figure 5C). Moreover, regular bimorph actuators need higher voltage values to initiate thermal actuation (onset voltage = 4V). On the other hand, molecular-composite bimorph actuators exhibit much superior curvature values of 0.22 cm$^{-1}$ for GO/TR42 and 0.51 for GO/TR15 with the same input power (Figure 5C). This reflects the significantly higher energy efficiency ($\eta \propto$ curvature$^2$) of the molecular-composite bimorph actuators with respect to regular composite bimorph actuators. [29] Molecular-composite bimorph actuators made of GO/TR42 can reach drastic curvature values (1.2 cm$^{-1}$) at a relatively small length (1.6 cm); this behavior potentially originates from the superior mechanical strength and elongation at break values of TR42 (Figure 5C-inset).

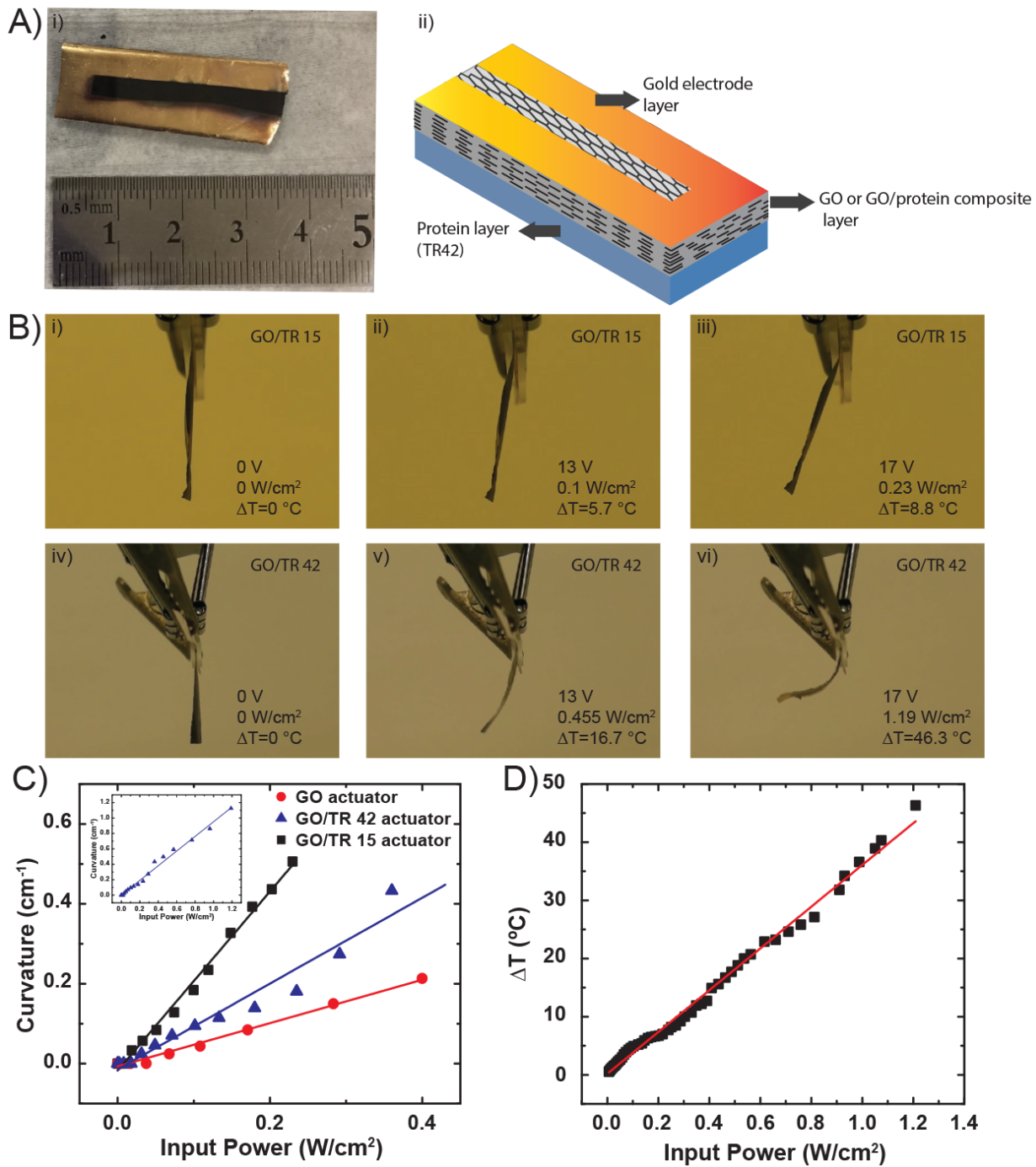

**Figure 5.** A) i) Image and ii) schematic illustration of thermal actuators. Images of the thermal actuators fabricated using GO/TR 15 (i-iii) and GO/TR42 (iv-vi) under different voltages. C) Curvature values of thermal actuators fabricated using GO, GO/TR15 and GO/TR42 films as a function of the applied power. D) Relative change in temperature as a function of the applied power.

The high power values (1.2 W/cm$^2$) required to achieve this actuation are rather deceiving and potentially originate from the low heating efficiency of the electrodes (Figure 5C-inset). The change in temperature needed to establish a curvature of 1.2 cm$^{-1}$ is only 47 ºC for this actuator, which is comparable to the state of the art thermal actuators (Figure 5D).[30, 31] The onset voltage of actuation for bimorph actuators consisting of GO/TR42 (3V) is slightly lower than regular bimorph actuators, due to their higher energy efficiency. In contrast to molecular composites prepared with higher number of tandem repeats (i.e. high molecular weight), molecular composites with lower-molecular-weight TR proteins enhances the energy efficiency further, since they can reach higher curvature values with lower input power for thermal actuation (Figure 5C). These results are consistent with Timoshenko beam theory, as initially proposed for the analysis of bi-metal thermostats [32] (i.e., the interlayer thickness of TR layer inversely scales with the curvature). Consequently, the onset voltage of actuation for GO/TR15 bimorph actuators (2V) is the lowest among the regular and molecular-composite bimorph actuators. In addition, these actuators can be relaxed rapidly to their initial state ($t_{relaxation}$<2s) by immersing them in water at room temperature, as the water acts as a plasticizer and renders these synthetic biopolymers rubbery.[21] Therefore, these highly efficient thermal actuators with rapid relaxation offer a better material alternative for thermal actuators in soft robotics and thermal sensing applications. Also, the low-temperature actuation (curvature: 1.2 cm$^{-1}$, T: 67 ºC) and biocompatible nature of the materials makes this actuation approach quite feasible for medical applications.

## 4. Conclusions

In summary, this novel 2D-layered hybrid material, in which interlayer distances can be precisely tuned by the molecular weight of the protein component, presents a very novel platform to study how thermal actuation can be controlled using molecular composites. This molecular-composite approach offers control over two critical parameters for thermal actuation, flexibility (maximum achievable curvature) and energy efficiency (curvature/input power). In this study, we demonstrated that curvature of the bimorph GO actuators can be improved up to 1.2 cm$^{-1}$ by increasing the number of repeats (i.e. molecular weight) of TR proteins; this curvature is 3 times higher than that possible with conventional GO/TR composites. In addition, bimorph GO actuators fabricated using molecular composites exhibited energy efficiencies nearly 4 times higher than regular bimorph GO actuators. This work also revealed that the energy efficiency of actuation scales inversely with the number of protein tandem repeats and consequently the interlayer distance of the GO sheets. Therefore, it is possible to further improve the energy efficiency of bimorph GO/TR actuators by decreasing the number of tandem repeats of these synthetic proteins. The molecular-composite actuators facilitated thermal actuation at activation voltages as low as 2V, and energy efficiencies approximately 18 times higher than the regular bimorph actuators; these nanoscale characteristics are important to the operation of flexible 2D devices made from these materials. This work presents the initial groundwork and the potential for using tandem-repeat structural proteins in composite with 2D-layered materials (beyond graphene [33]) to initiate more efficient materials for communication, sensing, repair and autonomy.

## Acknowledgments

MCD, BDA, AP, and HJ were supported partially by the Army Research Office under grant No. W911NF-16-1-0019, and Materials Research Institute of the Pennsylvania State University. MT and YL were supported by the Army Research Office under MURI grant No. W911NF-11-1-0362. We thank Dr. Jennifer Gray for helping in obtaining TEM images in this manuscript.

## Author contributions

MCD and MV conceived the idea. YL worked on GO preparation under the supervision of MT. MV fabricated the composite protein-GO films and performed actuator measurements. AP characterized molecular composites using spectroscopic, structural and thermal techniques. HJ worked on the cloning, recombinant expression and purification of proteins together with BDA. All authors contributed to writing and revising the manuscript, and agreed on the final content of the manuscript.

## Conflict of interest statement

The authors have a pending patent application on the concept of 2D-layered molecular composites.